\documentclass[universe,article,accept,moreauthors,12pt,a4paper]{mdpi} 

\usepackage{graphicx}
\graphicspath{{./}}

\makeatletter
\g@addto@macro{\UrlBreaks}{\UrlOrds}
\makeatother

\usepackage[utf8]{inputenc}
\usepackage[english]{babel}
\usepackage{amsmath}
\usepackage{amsfonts}
\usepackage{amssymb}
\usepackage{epsfig}
\usepackage{graphics,psfrag,rotating}
\usepackage{graphicx}
\usepackage{dcolumn}
\usepackage{bm}
\usepackage{textcomp}
\usepackage{keyval}
\usepackage{epstopdf}
\usepackage{color}
\usepackage[usenames,dvipsnames,svgnames]{xcolor}
\usepackage[T1]{fontenc}
\usepackage{multirow}
\usepackage{float}
\usepackage{subfigure}
\usepackage{xcolor}

\lastpage{x}
\doinum{10.3390/------}
\pubvolume{2}
\pubyear{2016}
\externaleditor{Academic Editors:  Mariusz P. Dąbrowski, Manuel Krämer and Vincenzo Salzano}
\history{Received: 30 October 2016; Accepted: 12 December 2016; Published: date}

\Title{New Constraints on Spatial Variations of the Fine Structure
Constant from Clusters of Galaxies}

\Author{ Ivan De Martino $^{1,}$*, Carlos J. A. P. Martins $^{2,3}$, Harald Ebeling $^{4}$ and Dale Kocevski $^{5}$}

\address{%
$^{1}$ \quad Faculty of Science and Technology, Department of Theoretical Physics and History of Science,
University~of~the~Basque Country (UPV/EHU), Barrio Sarriena s/n, 48940 Leioa, Spain\\
$^{2}$ \quad Centro de Astrofisica da Universidade do Porto, Rua das Estrelas s/n, 4150-762 Porto, Portugal; Carlos.Martins@astro.up.pt\\
$^{3}$ \quad Instituto de Astrof\'{\i}sica e Ci\^encias do Espa\c co, CAUP, 
Rua das Estrelas, 4150-762 Porto, Portuga.\\
$^{4}$ \quad Institute for Astronomy, University of Hawaii, Honolulu, HI 96822, USA; ebeling@ifa.hawaii.edu\\
$^{5}$ \quad Department of Physics and Astronomy, Colby College, Waterville, ME 04901, USA; dale.kocevski@colby.edu}

\corres{Correspondence: ivan.demartino1983@gmail.com}

\abstract{We have constrained the spatial variation
of the fine structure constant using multi-frequency measurements of the thermal Sunyaev-Zeldovich effect
of 618 X-ray selected clusters. Although our results are not competitive with the ones from 
quasar absorption lines, we improved by a factor 10 and $\sim$2.5  previous results from Cosmic Microwave Background
power spectrum and from galaxy clusters, respectively.}

\begin{document}

\section{Introduction}\label{sec:intro}

Recently, large observational studies of Quasar (QSO) absorption lines claimed
that the fine structure constant could vary over the sky with a dipolar modulation 
\cite{webb2011, King2012, Berengut2011, Berengut2012,Pinho2016}. The direction of such a dipole is comparable 
to the direction of other so-called anomalies and/or dipoles \cite{Mariano2012, Mariano2013}. However,
complementary analyses carried out on the Cosmic Microwave Background (CMB) power spectrum \cite{planck_int_24, bryan2015},
and studying the effect of such modulation on the orbital motion of the major bodies of the solar system \cite{iorio2011},
do not report any detection. 
Since their sensitivity is, however, one thousand times worse than the one achieved with 
QSO data, there is a need to carry out tests with other observational datasets 
such as multi-frequency measurements of the thermal Sunyaev Zeldovich effect (TSZ, \cite{tsz}) 
in cluster of galaxies. 

The TSZ effect induces secondary anisotropies on the CMB power spectrum. Such anisotropies are 
usually expressed in terms of the Comptonization parameter ($Y_c$) as $\Delta T_{TSZ}=T_0G(\tilde{\nu})Y_c$,
where $G(\tilde{\nu})$ is the TSZ spectral dependence, $Y_c$ is proportional to the integral  along the line of sight 
of the pressure profile of the intra-cluster medium, and $T_0$ is the current value of the CMB black-body temperature.
Moreover, 
$\tilde{\nu}=h\nu(z)/k_BT_{CMB}(z)$ is the reduced frequency, $k_B$ is the Boltzmann constant,
$h$ is the Planck constant and $T_{CMB}(z)$ is the CMB black-body temperature at the cluster location.

In the concordance $\Lambda$CDM
 model, the CMB black-body temperature is 
$T_{CMB, std}(z)=T_0(1+z)$. Any departure from such behavior should imply an extension of the concordance
model. In particular classes of models, where  a scalar field is coupled 
to the Maxwell $F^2$ term in the matter Lagrangian, the~photon number conservation is violated. 
Thus, the violation of the standard $T_{CMB}(z)$ law is coupled to the variation of the fine-structure as
\begin{equation}\label{eq:tcmb}
\frac{T_{CMB}(z)}{T_0}\sim(1+z)\left(1+\epsilon\frac{\Delta\alpha}{\alpha}\right)\,
\end{equation}
and their common origin implies that the coefficient $\epsilon$ is expected to be of order unity \cite{Avgoustidis2014}; for example, in a somewhat simplistic adiabatic limit $\epsilon=1/4$. 
Thus, if one is able to determine $T_{CMB}(z)$ at the cluster location using multi-frequency
measurements of the TSZ effect, then Equation \eqref{eq:tcmb} 
can be used as a phenomenological relation to observationally test the spatial variation  of $\alpha$. 

We apply the techniques developed in \cite{demartino2015}
to measure the TSZ emission at the location of 618 X-ray selected clusters. Then, we use
such measurements to test the spatial variation of the fine structure constant.  
The article is organized as follows: in Section \ref{sec:data} we will describe the data; 
in~Section~\ref{sec3}  we will explain the methodology; in Section \ref{sec4}  we will illustrate and discuss our results; 
and finally, in~Section~\ref{sec5}, we will give our main conclusions.

\section{Data} \label{sec:data}

Galaxy clusters were selected from the ROSAT-ESO Flux Limited X-ray catalog (REFLEX, \cite{bohringer04}), 
the extended Brightest Cluster Sample (eBCS, \cite{ebeling98, ebeling00})
and the Clusters in the Zone of Avoidance (CIZA,~\cite{ebeling02}). 
The catalog lists 618 galaxy clusters for which all quantities of interest for our analysis are directly measured or derived:
positions, spectroscopic redshifts, the central Comptonization parameter ($y_{c,0}$), 
the scale radius at which the mean overdensity of
the cluster is 500 times the critical density ($r_{500}$), and 
the corresponding angular size $\theta_{500}=r_{500}/d_A(z)$. 
Here, $d_A(z)$ is the angular diameter distance of each cluster.   
At the location of each cluster we measure the TSZ anisotropies using the {\it Planck} 2013 maps 
(Maps were originally released in a Healpix format with resolution $N_{side} = 2048$ \cite{gorski2005}. 
Data are available at http://www.cosmos.esa.int/web/planck).

In March 2013, the Planck Collaboration made publicly available its nine 
Nominal maps covering the frequency channels from 30 to 857 GHz. Since the TSZ effect has a peculiar spectral dependence 
it can be distinguished from other components and reliably detected once 
the foreground emissions  and the cosmological CMB signal are reduced.
For that purpose, we apply a cleaning procedure, fully described in \cite{demartino2015, demartino2016a, demartino2016b},
to the High Frequency Instrument data (100--857 GHz), since they have better resolution and lower instrumental noise 
than the Low Frequency Instrument data (30--70 GHz). 
{\it Before applying the cleaning procedure, we remove the intrinsic CMB monopole and dipole from the Nominal
 maps} (We use the remove\_dipole.pro facility of Healpix. It simultaneously performs a least square 
 fit of the monopole and dipole on all the valid pixels and removes them from the map. More details can be found at
  http://healpix.jpl.nasa.gov/html/subroutinesnode86.htm).  
The cleaning procedure returns, for each Planck frequency channel,  a foreground cleaned patch centered at the position 
of each cluster in our catalog. 
Then, in each patch and for each frequency, we measure the average TSZ temperature ($\delta \bar{T}/T_0 (\nu)$) 
over discs with angular extent equal to $\theta_{500}$. Hereafter, 
the averaged quantities are always evaluated on discs of radius $\theta_{500}$ and indicated with
a bar. To assign an error bar to each of the measurements, we carry out 1000 random simulations. In each one, we 
evaluate the mean temperature fluctuations in patches randomly placed 
out of the cluster positions and clean using the same procedure adopted for 
the real cluster population. Finally, we compute the correlation matrix between 
different frequencies ($C_{ij}$)
 \begin{equation}
 C_{ij}\equiv C(\nu_i,\nu_j)=\frac{\langle [\delta \bar{T}(\nu_i)-\mu(\nu_i)]
 [\delta  \bar{T}(\nu_j)-\mu(\nu_j)]\rangle}{\sigma(\nu_i)\sigma(\nu_j)} ,
 \label{eq:Cij}
 \end{equation}
 {where the average was computed over all simulations, $\mu(\nu_i)=\langle\delta  \bar{T}(\nu_i)\rangle$, and 
 \mbox{$\sigma(\nu_i)=\langle [\delta  \bar{T}(\nu_i) -\mu(\nu_i)]^2\rangle^{1/2}$}.} The square roots of 
the diagonal elements of the correlation matrix are the error bars associated to the TSZ temperature anisotropies of each cluster. 

\section{Methodology} \label{sec3}

To estimate the CMB temperature at the cluster location, as well as to
constrain the spatial variations of the fine structure constant, we 
use a Monte Carlo Markov Chain (MCMC) technique to explore the 
corresponding  relevant parameter space. Our pipelines 
employ  the Metropolis-Hastings sampling algorithm \cite{Metropolis1953, Hastings1970} 
with different (randomly set) starting points, and the Gelman-Rubin criteria~\citep{Gelman1992}
to test the convergence and the mixing of each chain. Specifically, 
we compute the ratio of the variances in the
target distribution, and require it to be $\mathcal{R} < 1.03$ 
(mathematical definitions are given in Section 3.2 of \cite{Verde2003}).  
We run four chains that stop when they satisfy the  criteria and contain at least 30,000 steps.
Finally the chains are merged, and the expectation
value of the 1D marginalized likelihood distribution is computed using all the steps.

First we need to estimate the CMB temperature at the cluster location. Having measured for each 
galaxy cluster the averaged TSZ emission as a function of frequency, we predict the theoretical 
counterpart at the same aperture $
\Delta \bar{T}({\bf p}, \nu_i)/T_0 =G(\nu_i, T_{CMB}(z)) \bar{Y}_c$, 
where ${\bf p}=[T_{CMB}(z),\bar{Y}_c]$ are the free parameters of the model, 
and $\bar{Y}_c$ is  the averaged Comptonization parameter. Finally, we compute 
the likelihood $-2\log{\cal L}=\chi^2({ \bf p})$ as
\begin{equation}
-2\log{\cal L}=\chi^2 ({ \bf p})=\Sigma_{i=0}^{N} \Sigma_{j=0}^{N} 
\Delta \bar{T}_{i}({\bf p})C_{ij}^{-1}
\Delta \bar{T}_{j}({\bf p}),
\label{eq:chi}
\end{equation}
where $\Delta \bar{T}_{i}({\bf p})\equiv\dfrac{\Delta \bar{T}({\bf p}, 
\nu_i)}{T_0}-\dfrac{\delta \bar{T}(\nu_i)}{T_0}$, and
$N=4$ is the number of data points (the four frequencies).
Once our simulations reached the convergence criteria, we estimated the CMB temperature
for each individual cluster, and its 68\% of confidence level error bar, from
the 1D marginalized likelihood distribution.
See \cite{demartino2016c} for a full description of the methodology.
These measurements are used to estimate the $\alpha$-anisotropies 
as 
\begin{equation}
 \frac{\Delta\alpha}{\alpha}=\epsilon^{-1}\left(1-\frac{T_{CMB}(z)}{T_0(1+z)}\right).
\end{equation}

Finally, we probe 3 different models allowing for a monopole ($m$) amplitude plus dipole ($d$) variation.
Specifically, the first model 
assumes the functional form defined in \cite{webb2011} to fit quasar data ({\bf Model 1}) 
\begin{equation}\label{eq:model1}
 \frac{\Delta\alpha}{\alpha} = m + d \cos(\Theta),
\end{equation}
where  $\Theta$ is the angle on the sky between the line of sight of each cluster 
and the best fit dipole direction. Then, we fit the model used in \cite{Galli2013} to the cluster data.
Such a model ({\bf Model 2}) allows a variation with the $\Lambda$CDM look-back time, $r(z)=\int\frac{dz'}{H(z')}$:
\begin{equation}\label{eq:model2}
 \frac{\Delta\alpha}{\alpha} = m + d r(z)\cos(\Theta).
\end{equation}
 
 Finally, to explore the possibility of an intrinsic dipole, we also try to fit the following relation to the CMB temperature measurements 
  ({\bf Model 3})
\begin{equation}\label{eq:model3}
\frac{\Delta T}{T_0(1+z)} = m + d \cos(\Theta).
\end{equation} 

For each model we analyze four different configurations of the parameter space: (A) $m=0$ for the Model 1 and 2, and unity for Model 3,
the direction of dipole is fixed to the one from QSO \cite{webb2011}, while the amplitude vary in the interval $[-1,1]$;
(B) the direction is still fixed, but we also vary the monopole in the intervals $[-1,1]$ for Model 1 and 2, and $[0,2]$ for Model 3;
(C) and (D) we repeat the configurations of (A) and (B) but we also vary the dipole direction over the whole sky.

\section{Results and Discussion}
\label{sec4}

\begin{figure}[H]
 \centering
 \includegraphics[width=0.95\columnwidth]{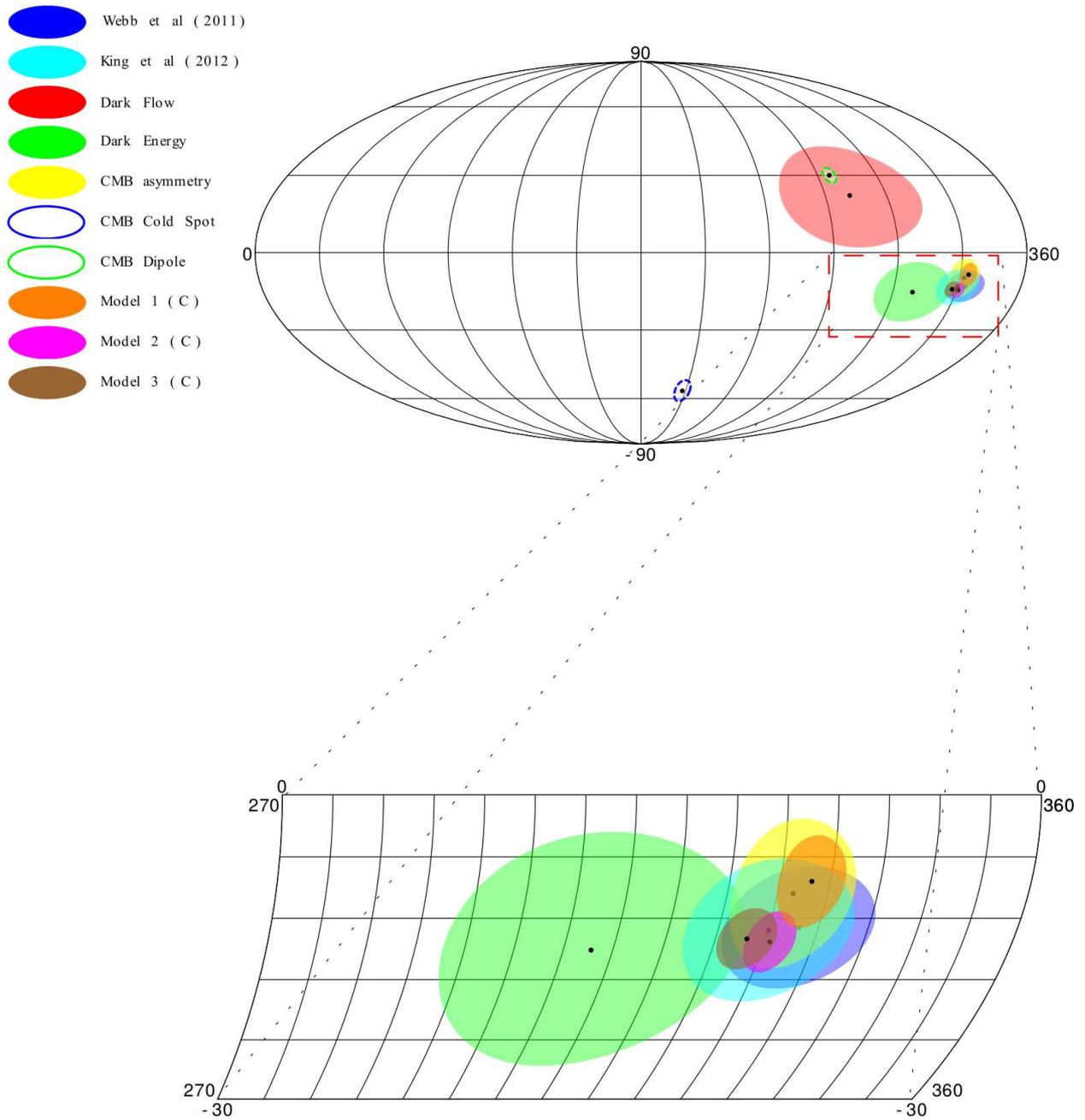}
\caption{Directions in galactic coordinates for the spatial variation of the fine structure constant 
 from~\cite{webb2011,King2012} in blue and cyan, respectively, for the Dark Energy dipoles from 
 \cite{Mariano2012} in green, Dark Flow direction from \cite{darkflow2008, darkflow2010, darkflow2011, darkflow2012, darkflow2015} in red,
for the Cosmic Microwave Background (CMB) asymmetry  from {\cite{Mariano2013}} in yellow, and for our results from the analysis (C) of models 1,2 and 3 in
orange, magenta and brown, respectively. Finally, we also indicate in blue and green dashed circles the direction of the 
Cold Spot anomaly and the intrinsic CMB dipole, respectively \cite{Vielva2004, planck13_XXIII,Kogut1993}.}\label{fig5}
\end{figure}

{ Our results show that the monopole and dipole amplitudes are always compatible with their expected valued in standard cosmology
within 1.5$\sigma$. The best constraints are: $m=0.006\pm0.004$  and $d=-0.008\pm0.009$ for Model 1 (B),
and $d=-0.003\pm0.003$ GLyr$^{-1}$ and  $m=0.006\pm0.005$ for Model 2 (A) and (B), respectively.
The direction of the ``possible'' dipole is always compatible with the one predicted from QSO in \cite{webb2011}, but when 
they are varied the constraints on the monopole and dipole amplitudes are degraded since the number of free parameters is increased.
 Nevertheless, our best fit directions are always compatible
with the one from QSO data \cite{webb2011, King2012} and other CMB anomalies~\cite{Mariano2012, Mariano2013},
while they are several degrees away  from the directions of the intrinsic CMB and the Dark Flow dipoles, 
from the kinetic CMB asymmetry, and from the CMB cold spot location \cite{Kogut1993, darkflow2008, darkflow2010, darkflow2011, darkflow2012, darkflow2015, 
Vielva2004, planck13_XXIII, Zhao2012, Zhao2013, Zhao2014, Zhao2015, Zhao2016}. 
All results are summarized in the Table \ref{tab:results} and, for easier comparison, the anomalies directions 
are plotted in Figure \ref{fig5}.}

Since our analysis shows that the possible dipole direction coincides with the one  obtained from QSO \cite{webb2011}
and dark energy models \cite{Mariano2012,Mariano2013}, it could be interpreted as the fact that such dipoles, if confirmed 
with more extended and independent analysis, could represent signature of the failure of the 
$\Lambda$CDM model in favor of varying constant models or extended theories of gravity.
Finally let us remark that, although we are not competitive with the QSO constraint from \cite{webb2011}, our best fits for 
Model 2 provide a precision improvement of a factor $\sim$2.5, while being compatible with the results in \cite{Galli2013}. 
For Model 1, we improve on the previous constraints in \cite{planck_int_24, bryan2015} by a factor $\sim$10. 
The improvement is mainly due to the fact that previous analyses, based on the CMB power spectrum, need 
to fit not only all "standard" parameters of the cosmological model, but also the additional ones related to the
variation of fundamental constants; as a consequence, the resulting constraints were less tight.

\unskip

\section{Conclusions}
\label{sec5}

We have considered a model in which the variation of the fine structure constant is coupled
to the variation of the CMB temperature. Using Equation \eqref{eq:tcmb}, we have
constrained the spatial variation of the fine structure constant using the Planck 2013 Nominal maps 
and a X-ray selected cluster catalog. First, we have cleaned the Planck maps and extracted 
the CMB temperature at the location of 618 X-ray selected clusters using the TSZ multi-frequency measurements.
Then, we have carried out a statistical analysis to test three models allowing for 
monopole and dipole amplitudes, and describing both spatial variations of $\Delta \alpha/\alpha$ and of $T_{CMB}$ itself. All results of our analysis are summarized in Table~\ref{tab:results}.

\begin{table}[H]
\centering
\small
\begin{tabular}{cccccc}
\hline
 &  &    \textbf{\emph{m}} & \textbf{\emph{d}} & \boldmath\textbf{RA ($^\circ$)} & \boldmath\textbf{DEC ($^\circ$)} \\
 \hline
\multirow{4}{*}{Model 1} & (A) &  0.0               & $-0.002\pm0.008$   &      261.0         &  $-58.0$   \\
                        & (B) &  $0.006\pm0.004$   & $-0.008\pm0.009$   &      261.0         &  $-58.0$        \\
                        & (C) &     0.0              & $-0.030\pm0.020$    &     $255.1\pm3.8$  &  $-63.2\pm2.6$  \\
                        & (D) &  $0.021\pm0.029$   & $-0.030\pm0.014$    &     $255.9\pm4.2$ &  $-55.3\pm5.8$   \\
 \hline
\multirow{4}{*}{Model 2} & (A) &   0.0          & $-0.003\pm0.003$ GLyr$^{-1}$   &      261.0          &  $-58.0$     \\
                        & (B) &  $0.006\pm0.005$   & $-0.003\pm0.005$ GLyr$^{-1}$ &      261.0          &  $-58.0$       \\
                        & (C) &      0.0           & $-0.042\pm0.049$ GLyr$^{-1}$   &      $261.6\pm16.1$ &  $-61.3\pm2.7$   \\
                        & (D) &  $0.019\pm0.011$   & $-0.027\pm0.051$ GLyr$^{-1}$   &      $245.0\pm12.9$ &  $-56.0\pm3.8$   \\ 
\hline
\multirow{4}{*}{Model 3} & (A) &  1.0                 & $-0.010\pm0.008$     &      261.0         &  $-58.0$  \\
                        & (B) &  $1.001\pm0.002$   & $-0.003\pm0.002$    &      261.0         &  $-58.0$       \\
                        & (C) &     1.0            & $-0.020\pm0.015$     &      $258.0\pm1.2$ &  $-64.0\pm1.1$  \\
                        & (D) &  $1.000\pm0.0001$  & $-0.018\pm0.015$    &      $258.4\pm1.9$ &  $-64.3\pm2.6$   \\
 \hline
\end{tabular}
\caption{Results of the analysis for Models 1, 2 and 3 from \cite{demartino2016c}.}\label{tab:results}
\end{table}

Although at the present time galaxy clusters are not competitive with high-resolution spectroscopic measurements in
absorption systems along the line-of-sight of bright quasars, our~analysis was improved by an order of magnitude previous constraints by the Planck Collaboration \cite{planck_int_24} and by a factor $\sim$2.5 the results from Galli (2013) \cite{Galli2013}.

\vspace{6pt}
\section*{Acknowledgments}
I.D.M. acknowledges the financial support from the University of the Basque Country UPV/EHU under the program
``Convocatoria de contrataci\'{o}n para la especializaci\'{o}n de personal 
investigador doctor en la UPV/EHU 2015'', from the Spanish Ministry of 
Economy and Competitiveness through research project FIS2014-57956-P (comprising FEDER funds)
and from the Basque Government grant IT956-16 for the GIC research group.
This article is based upon work from COST Action CA1511 Cosmology and Astrophysics 
Network for Theoretical Advances and Training Actions (CANTATA), 
supported by COST (European Cooperation in Science and Technology).
This research was conducted in the context of project PTDC/FIS/111725/2009 (FCT, Portugal), 
with additional support from grant UID/FIS/04434/2013. C.J.M. is also supported by an 
FCT Research Professorship, contract reference IF/00064/2012, funded by FCT/MCTES 
(Portugal) and POPH/FSE (EC). C.J.M. thanks the Galileo Galilei Institute for Theoretical 
Physics for the hospitality and the INFN for partial support during the completion of this work.

\section*{Author contributions}IDM and CJM conceived the project and wrote the paper; IDM performed the
data analysis; HE and DK provided the X-ray cluster catalog.

\conflictofinterests{
The authors declare no conflict of interest. }

\bibliographystyle{mdpi}
\renewcommand\bibname{References}

\end{document}